\documentclass[letterpaper, 10 pt, conference]{ieeeconf}  

\IEEEoverridecommandlockouts                              

\usepackage{graphicx} 
\usepackage{amsmath} 
\usepackage{amssymb}  
\usepackage{multirow}

\title{\LARGE \bf
Low-rank Similarity Measure for Role Model Extraction*
\thanks{*This paper presents research results of the Belgian Network DYSCO (Dynamical Systems, Control, and Optimization), funded by the Interuniversity Attraction Poles Programme, initiated by the Belgian State, Science Policy Office. The scientific responsibility rests with its author(s).}
}

\author{Arnaud Browet$^\dagger$ and Paul Van Dooren$^\dagger$ 
\thanks{$^\dagger$Institute of Information and Communication Technologies, Electronics and Applied Mathematics (ICTEAM), Catholic University of Louvain (UCL), 4 Av. G. Lemaitre,	1348 Louvain-la-Neuve, Belgium. }
\thanks{Corresponding author: {\tt\small arnaud.browet@uclouvain.be}}
}

\newcommand{\Rnn}{\mathbb{R}^{n\times n}}
\newcommand{\Rnr}{\mathbb{R}^{n\times r}}
\newcommand{\Rrr}{\mathbb{R}^{r\times r}}
\newcommand{\RnNnN}{\mathbb{R}^{n^2\times n^2}}
\newcommand{\pipe}{\;\left|\right.}
\newcommand{\bigo}{O}
\newcommand{\eref}[1]{(\ref{#1})}
\newcommand{\fref}[1]{Fig. \ref{#1}}
\newcommand{\ie}{i.e.\ }
\newcommand{\etal}{et al.\ }
\newcommand{\eg}{e.g.\ }

\newcommand{\norm}[1]{\left\|#1\right\|}

\begin{document}

\maketitle
\thispagestyle{empty}
\pagestyle{empty}

\begin{abstract}
Computing meaningful clusters of nodes is crucial to analyze large networks. In this paper, we present a pairwise node similarity measure that allows to extract roles, \ie group of nodes sharing similar flow patterns within a network. We propose a low rank iterative scheme to approximate the similarity measure for very large networks. Finally, we show that our low rank similarity score successfully extracts the different roles in random graphs and that its performances are similar to the full rank pairwise similarity measure.
\end{abstract}

\section{INTRODUCTION}

Many complex systems might be represented as network structures, for example, human interactions or mobile phone telecommunications, food webs or gene interactions. In recent works, a lot of attention has been focused on the extraction of meaningful clusters to characterize networks at different levels \cite{Arenas2008}. This clustering is essential to comprehend large networks and extract relevant statistical properties. Many researchers have proposed appropriate measures and algorithms to unfold community structures, i.e. groups of densely connected nodes \cite{Porter2009, Fortunato2010}. However, this structural distribution of nodes in networks is not always representative and lack generalization in practical contexts. For instance, bipartite networks or cycle graphs do not contain communities although they may be heavily structured. Less attention has been paid to uncover more general structures which is known as roles extraction or block modeling \cite{Wasserman1994,Cason2012}. In previous work \cite{reichardt2007}, Reichardt \& White had applied a similar approach than community detection in the framework of \cite{reichardt2006} to extract roles in networks. In this paper, we assume that the different roles in a network should represent groups of nodes sharing the same behavior within the graph or, in other words, having similar flow patterns. This generalized the notion of communities which can also be described as roles where each node in a role mainly interacts with other nodes in the same role. But many other role interactions may be defined like, for example, a leader-follower model on social network interactions or a block cycle model for food webs. In this paper, we present a pairwise node similarity measure designed to derive such role models. This similarity measure compares the neighborhood patterns of every node and is expected to be high for any pair of nodes sharing analogous flow properties. Since computing the exact pairwise similarity is computationally expensive, we propose a low rank iterative scheme that approximates the similarity score and allows to analyze large networks. We will first present the similarity measure defined as the fixed point solution of a converging sequence. We will then introduce our low rank approximation and briefly demonstrate its convergence. Finally, we will apply the similarity measure and our low rank approximation to random graphs containing a structural block distribution of nodes, and show that they successfully extract the different roles within this kind of graph. We will also exhibit some evidences that analyzing the evolution of the low rank similarity measure can reveal the number of roles in the network. Lastly, we will show that the performances of both measures are quantitatively equivalent hence justifying the application of our low rank iterative scheme in practical contexts.

\section{Node-to-Node similarity}
We consider a weighted and directed graph $G_A(V,E)$, with $V$ the set of vertices and $E$ the set of edges, associated to its adjacency matrix $A\in\Rnn$ where $A_{i,j}\neq0$ if $(i,j)\in E$ for $i,j \in V$. Our similarity measure should reveal nodes having similar behaviors in the network which we will identify by the neighborhood patterns of each node. We define a neighborhood pattern of length $\ell$ for a node as a sequence of length $\ell$  of incoming (I) and outgoing (O) edges starting from the node, which we will call the \emph{source} node. For example, the neighborhood patterns of length $1$ consist in exactly one edge and end up either in a parent (I) or in a child (O) of the source. If we consider neighborhood patterns of length $2$, then $4$ different types of nodes can be reached: the parent of a parent (I-I), the child of a parent (I-O), the parent of a child (O-I) or the child of a child (O-O). One can easily see that when the length of the neighborhood patterns is increased by $1$ the number of reachable nodes, which we will call the \emph{target} nodes, is doubled.

\begin{figure}[t]
	\centering
	\framebox{\parbox{0.85\columnwidth}{
		\includegraphics[width=0.85\columnwidth]{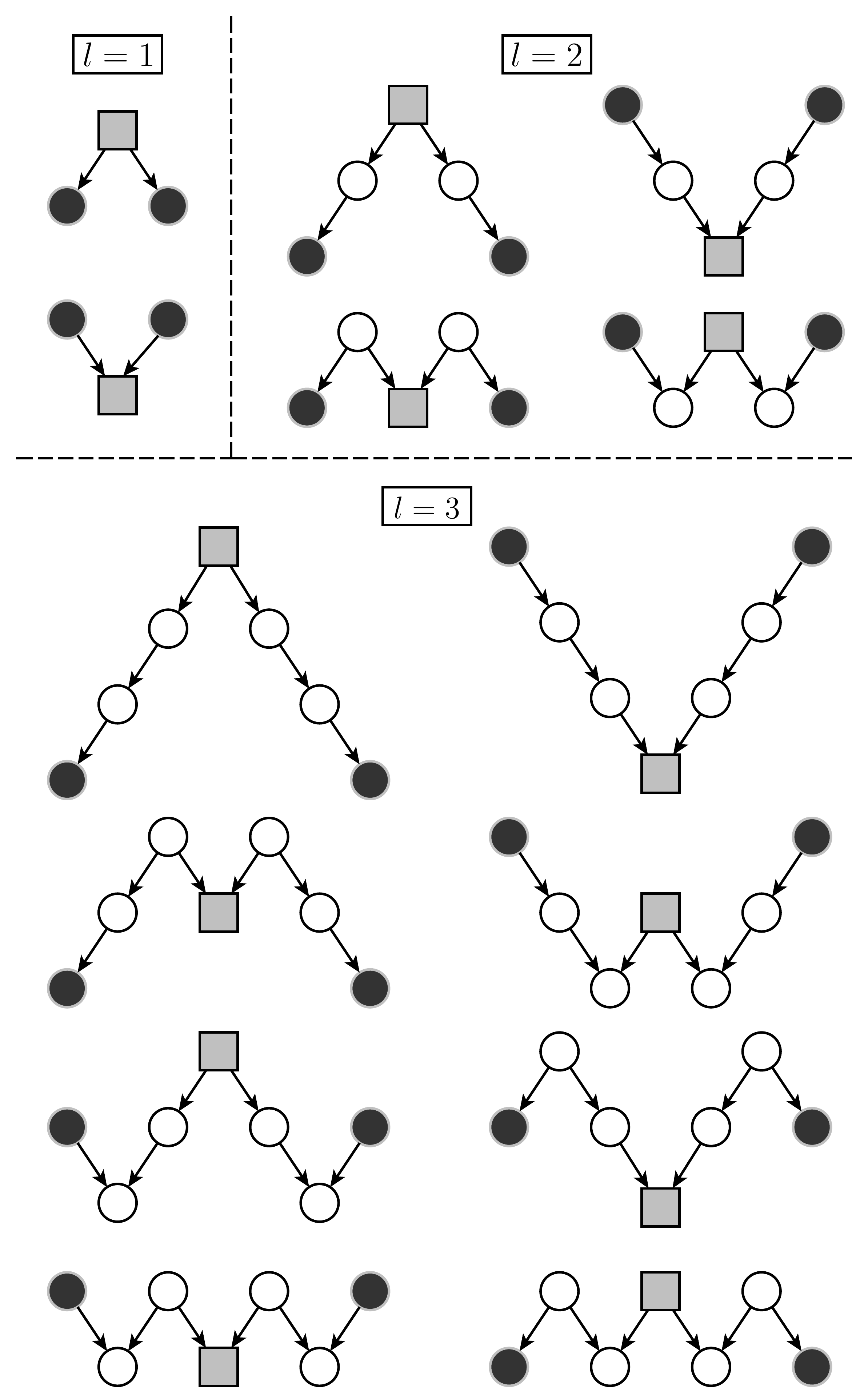}      
	}}
    \caption{All the different neighborhood patterns, up to length $3$, captured by the similarity measure $S_{i,j}$ \eref{eq:S} with the source nodes $i$ and $j$ represented as dark circles and the target node represented as a light gray square.}
	\label{fig:pattern}
\end{figure}

Our similarity measure reflects that a pair of nodes is highly similar if they have many neighborhood patterns in common, or in other words, if they can reach many targets with neighborhood patterns of the same kind and length. For example, using the patterns of length $1$, two source nodes will be more similar if they have many common parents (I) or many common children (O). \fref{fig:pattern} shows all the possible common neighborhood patterns, up to length $3$, where the source nodes are represented as dark circles and each target node as a light gray square. One can compute the number of common target nodes for every pair of source nodes using neighborhood patterns of length $1$ as $$N_1 = AA^T + A^TA,$$ where the first term gives the number of common children (O) and the second term gives the number of common parents (I). Similarly, the number of common target nodes for neighborhood patterns of length $2$ is given by $$N_2 = AAA^TA^T + AA^TAA^T + A^TAA^TA + A^TA^TAA,$$
where the different terms corresponds to the neighborhood patterns (O-O), (O-I), (I-O) and (I-I), respectively.

Our pairwise node similarity measure $S\in\Rnn$, previously introduced in \cite{Denayer2012}, is then defined as
\begin{equation}
\label{eq:S_np}
S = \sum_{\ell=0}^{\infty} \beta^{2(\ell-1)}N_l,
\end{equation}
where $\beta\in\mathbb{R}$ is a scaling parameter. Hence our similarity measure computes the weighted sum of the number of common target nodes using neighborhood patterns of any length, and the contribution of the number of common targets using neighborhood patterns of length $\ell+1$ is weighted by $\beta^{2\ell}$. 

One can define an iterative sequence 
\begin{equation}
\label{eq:S}
S_{k+1} = \Gamma_A\left[I + \beta^2S_k\right],
\end{equation}
where $\Gamma$ is a linear operator,
$$\Gamma_A:\Rnn\rightarrow\Rnn:\Gamma_A[X] = AXA^T + A^TXA,$$
such that
$$S_{k+1} = \Gamma_A\left[I\right] + \dots + \left(\beta^{2}\right)^{k}\Gamma_A^{k+1}\left[I\right]+\left(\beta^2\right)^{k+1}\Gamma_A^{k+1}\left[S_0\right]$$
where $\Gamma_A^k[.]$ corresponds to applying $k$ times the operator $\Gamma_A$. Hence, our similarity measure $S$ can be computed as the fixed point solution of the iterative sequence \eref{eq:S}.

Our similarity measure $S$ \eref{eq:S_np} can be seen as a generalization of the measure proposed by {Cooper and Barahona} \cite{Cooper2011,Diaz2013} for which the pairwise similarity $S^{CB}$ only compares the total number of paths originating or leading to a node, without comparing the targets or the sources of those paths. Furthermore, this similarity score $S^{CB}$ does not consider all types of neighborhood patterns, as represented in \fref{fig:pattern}, but only restricts the measure to direct paths (represented in the first row of the panels $\ell=2$ and $\ell=3$ in the figure). While being easily computed, this makes the measure unable to extract a good pairwise similarity score for some particular graphs. For example, if one considers a regular block cycle graph, as represented in \fref{fig:cycle}, where each role contains the same number of nodes and each node is connected to all the nodes in the following role in the cycle, the pairwise similarity measure $S^{CB}$ is of rank $1$ because all the nodes have a constant number of in/out neighbors at all distances. This makes the extraction of roles in this network impossible using $S^{CB}$. On the contrary, our similarity measure \eref{eq:S} produces a fixed point solution $S^*$ of rank equal to the number of roles in the network, with an obvious clustering that reveals the different roles. One can see that any $2$ nodes of the same role in the input graph are isomorphic, while any $2$ nodes of different roles are not. This is accurately represented by our measure $S^*$ but not by $S^{CB}$.

The similarity measure we propose in this paper might also be compared to the self similarity score introduced by {Blondel} \etal\cite{Blondel2004}. However, this measure has some drawbacks that are avoided using our iterative scheme \eref{eq:S}, \ie the sequence $S_k$ converges for any initial matrix $S_0$ and the fixed point solution is unique. Moreover, it is known that the similarity score of {Blondel} \etal $S^B$ is of rank $1$ when the adjacency matrix $A$ is normal. After scaling, $S^{B}$ is therefore the matrix of all ones as $S^{CB}$, which makes the analysis of the block cycle graph again impossible using this similarity measure.

\def \pboxH {0.25\columnwidth}
\def \myframebox [#1] {{#1}}
\begin{figure}[t]
	\centering
	\framebox{\parbox[t][][c]{0.95\columnwidth}{
		\myframebox[\parbox[t][\pboxH][c]{0.2\columnwidth}
			{\includegraphics[width=0.2\columnwidth]{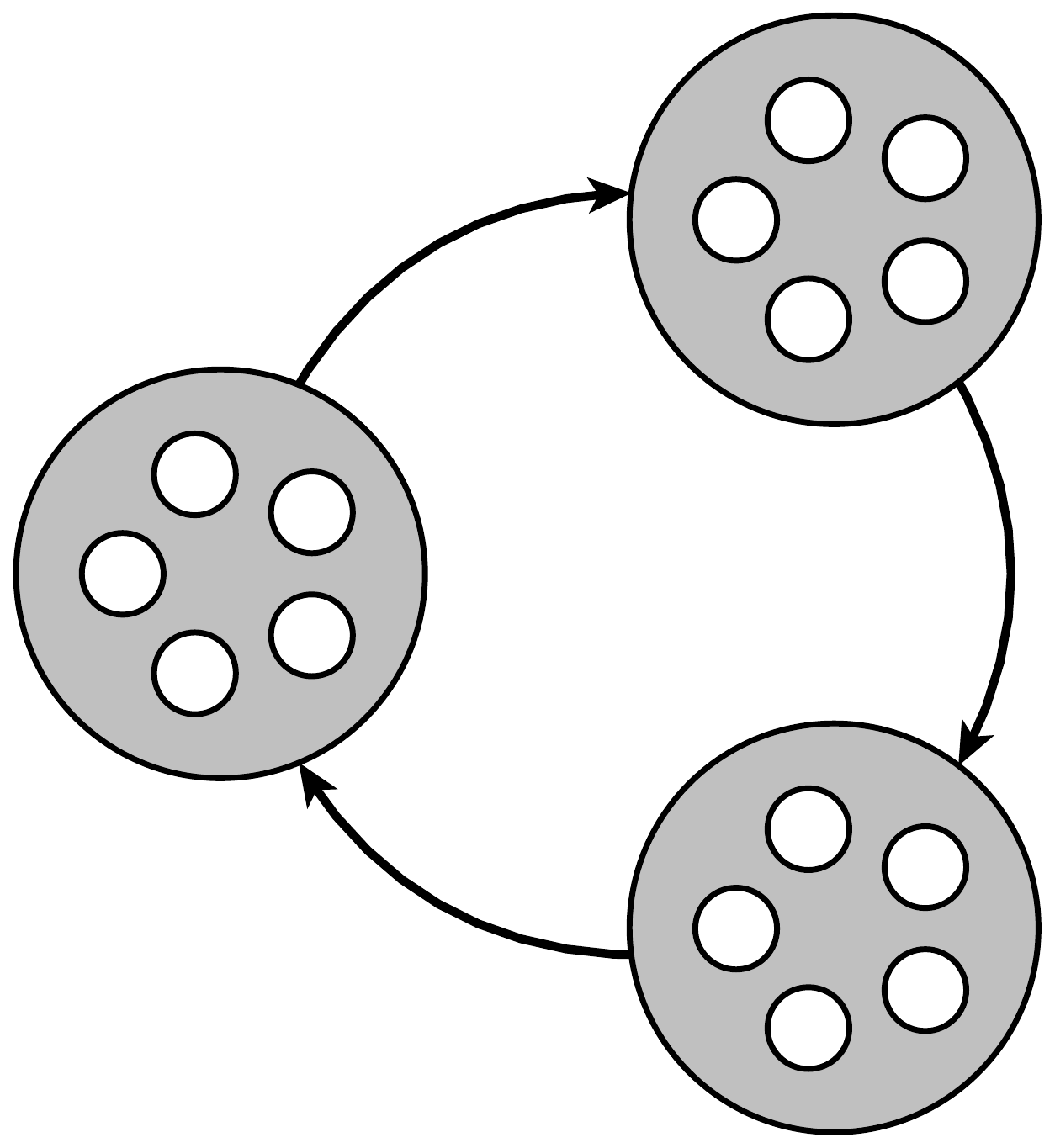}}]
		\hfill
		\myframebox[\parbox[t][\pboxH][c]{0.2\columnwidth}{\centering$A$\\[2pt]
			\includegraphics[width=0.2\columnwidth]{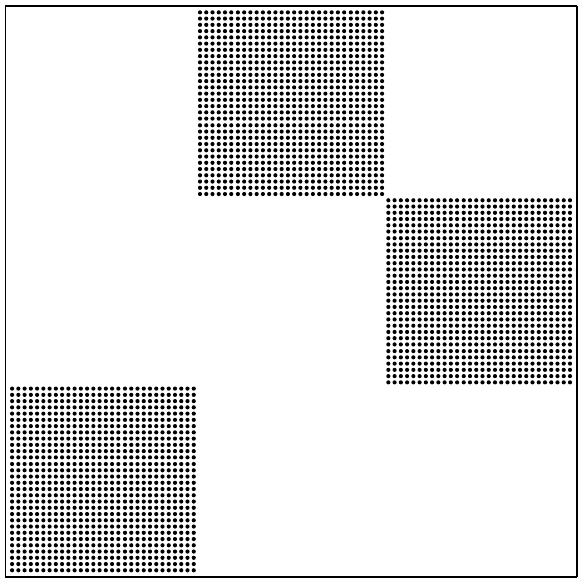}
			}]
		\hfill
		\myframebox[\parbox[t][\pboxH][c]{0.2\columnwidth}{\centering$S^*$\\[2pt]
			\includegraphics[width=0.2\columnwidth]{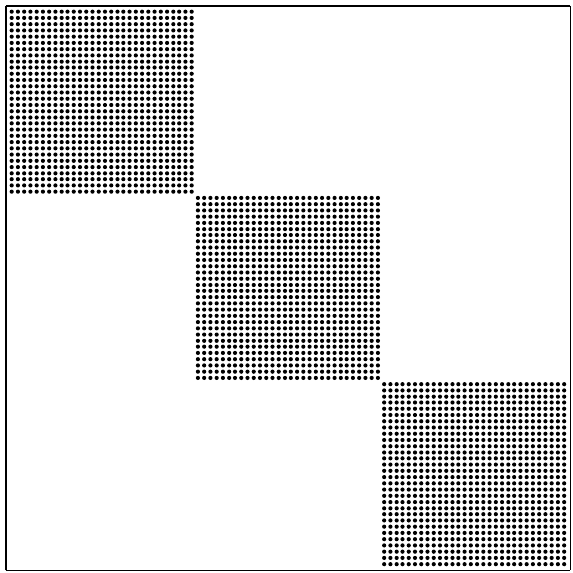}
			}]
		\hfill
		\myframebox[\parbox[t][\pboxH][c]{0.2\columnwidth}{\centering$S^{CB}=S^{B}$\\[2pt]
			\includegraphics[width=0.2\columnwidth]{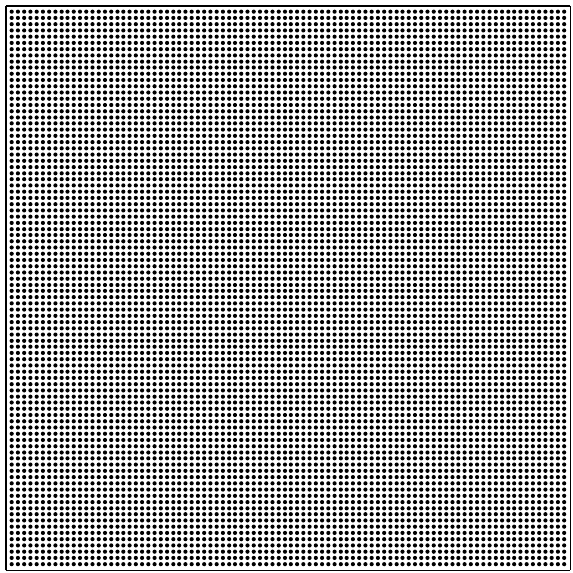}
			}]
	}}
    \caption{From left to right: Block cycle role graph where each block has the same number of nodes and each node is connected to all the nodes in the following block. The large gray filled circles represent the roles and the small white circles represent the nodes of the graph; The adjacency matrix of the block cycle graph; The fixed point pairwise similarity score $S^*$, computed using \eref{eq:S}, reveals all the different blocks; The pairwise similarity score of Cooper and Barahona $S^{CB}$ and Blondel \etal $S^B$ are rank $1$ and do not exhibit the block structure.}
	\label{fig:cycle}
\end{figure}

The parameter $\beta$ in \eref{eq:S} can be tuned to vary the weight of long neighborhood patterns but must be chosen wisely to ensure the convergence of the sequence $S_k$. If we initialize $S_0=0$, the iteration \eref{eq:S} can be written for $k\geq 1$ as 
\begin{equation}
S_{k+1} = S_1+ \beta^2\Gamma_A\left[S_k\right],
\label{eq:Ss}
\end{equation}
where 
\begin{equation}
S_1 = AA^T + A^TA,
\label{eq:S1}
\end{equation}
and the fixed point solution of \eref{eq:S} is then given by 
\begin{equation*}
S^* = S_1 + \beta^2\left(AS^*A^T + A^TS^*A\right),
\end{equation*}
if the sequence converges. Using a classical property of the Kronecker product, this can be written as 
\begin{equation*}
vec(S^*) = \left[I-\beta^2\left(A\otimes A + \left(A\otimes A\right)^T \right) \right]^{-1}vec\left(S_1\right)
\end{equation*}
where $vec(S)$ denotes the vectorization of the matrix $S$, formed by stacking the columns of $S$ into one single column vector. It follows that, to ensure convergence, one can choose $\beta$ such that
\begin{equation}
\beta^2\leq\frac{1}{\rho\left(A\otimes A + \left(A\otimes A\right)^T \right)}
\label{eq:beta}
\end{equation}
where $\rho(.)$ denotes the spectral radius. Computing the exact upper bound for the parameter $\beta$ to ensure convergence might be computationally expensive due to the Kronecker products $A\otimes A \in\RnNnN$ if $A$ is non-symmetric. However, one can use an easily computed bound
\begin{equation}
\beta^2 \leq \frac{1}{\rho\left((A+A^T)\otimes (A+A^T)\right)} = \frac{1}{\rho\left((A+A^T)\right)^2}
\label{eq:beta_bound}
\end{equation}
which ensures that the constraint \eref{eq:beta} is satisfied. However, even if $\beta$ is small enough to guarantee the convergence of the sequence \eref{eq:Ss}, it might be impossible to compute the fixed point solution up to a small tolerance because of the increasing computational cost and memory requirement. Indeed, even if $A$ is sparse, the matrix $S_k$ tends to  fill in as $k$ increases and each single iteration of \eref{eq:Ss} is $\bigo(n^3)$. This leads us to define a low-rank projected iteration to approximate the solution of \eref{eq:Ss}. In the next section, we will introduce the low-rank iteration and briefly demonstrate its convergence.

\section{Low-rank similarity approximation}
Because the full rank fixed point solution of \eref{eq:S} is often computationally too expensive to extract, we introduce a low-rank approximation of rank at most $r$ of $S^*$. Inspired from the formulation \eref{eq:Ss}, we define the low rank iterative scheme as
\begin{equation}
S^{(r)}_{k+1} = \Pi^{(r)}\left[ S^{(r)}_1+ \beta^2\Gamma_A\left[S^{(r)}_k\right]\right] = X_{k+1}\;X_{k+1}^{T}
\label{eq:LR}
\end{equation}
where $X_k \in \Rnr$ and $\Pi^{(r)}\left[.\right]$ is the best low-rank projection on the dominant subspace which can be computed using a truncated singular value decomposition (\textit{SVD}) of rank at most $r$. $S_1^{(r)}$ is the best low-rank approximation of $S_1$ which can be written as
$$S_1 = \left[A\pipe A^T\right]\left[A\pipe A^T\right]^T,$$
where $\left[A\pipe A^T\right]$ is the horizontal concatenation of $A$ and $A^T$. This allows us to efficiently compute  $S_1^{(r)}$ as
\begin{align*}
S_1^{(r)} &= \Pi^{(r)}\left[\left[A\pipe A^T\right]\left[A\pipe A^T\right]^T\right]\\
		&= U_1\Sigma_1^2U_1^T = X_1X_1^T
\end{align*}
where the columns of the unitary matrix $U_1\in\Rnr$ span the dominant subspace of dimension at most $r$ of $\left[A\pipe A^T\right]$ and $\Sigma_1\in\Rrr$ is a diagonal matrix of the dominant singular values, \ie $\left[A\pipe A^T\right] \approx U_1\Sigma_1V_1^T$. To compute each iterative solution of \eref{eq:LR}, one can see that
\begin{align*}
S^{(r)}_1+ \beta^2\Gamma_A\left[S^{(r)}_k\right] &= X_1X_1^T + \beta^2AX_kX_k^TA^T\\
&\hspace{25pt}+\beta^2A^TX_kX_k^TA\\
&=Y_k\;Y_k^T
\end{align*}
where
$$Y_k = \left[X_1\pipe\beta AX_k\pipe\beta A^TX_k\right],$$
which leads to
$$X_{k+1}X_{k+1}^T = \Pi^{(r)}\left[Y_kY_k^T\right].$$
To efficiently compute $X_{k+1}$, we first apply a \textit{QR} factorization to $Y_k = Q_kR_k$, then compute a truncated \textit{SVD} of rank at most $r$ of $R_k$ such that $R_k = \mathcal{U}_k{\Omega}_k\mathcal{V}_k$ and finally compute $$X_{k+1} = Q_k\mathcal{U}_k\Omega_k.$$
One can prove, using perturbation theory \cite{stewart1973}, that the iterative scheme \eref{eq:LR} converges locally to a fixed point solution $S^{(r)}$ if the spectral gap at the $r^{th}$ singular value is sufficiently large. Without going into the details of the demonstration of the convergence, let us mention some interesting results that follow from it. First, we consider the function $$f(S) = S_1^{(r)} + \beta^2\Gamma_A\left[S\right].$$
Clearly, since $S^{(r)}$ is a fixed point solution of \eref{eq:LR}, we know that there exist a unitary matrix $U\in\Rnr$ and a diagonal matrix $\Sigma\in\Rrr$ such that $S^{(r)} = U\Sigma^2U^T$ and
$$[U\;V]^T\;f(S^{(r)})\;[U\;V] = 
\left[\begin{array}{cc}
\Sigma^2 &\\
& \sigma^2
\end{array}\right]
$$
where $\Sigma_{i,i}>\sigma_{j,j}$ $\forall i,j$ because we assumed that the fixed point solution has a positive spectral gap at the $r^{th}$ singular value.

\noindent Then, we consider a small symmetric perturbation $\Delta$ and, using the linearity of the operator $\Gamma_A[.]$, one can write that $$f(S^{(r)}+\Delta) = f(S^{(r)}) + \beta^2\Gamma_A\left[\Delta\right]$$ and
$$
[U\;V]^T\;\left(f(S^{(r)}) + \beta^2\Gamma[\Delta]\right)\;[U\;V] = 
\left[\begin{array}{cc}
E_{11}&E_{21}^T\\
E_{21}&E_{22}
\end{array}\right].
$$
Since $U$ is in general not an invariant subspace of $f(S^{(r)}+\Delta)$, $E_{21}$ will be non-zero.
\noindent However, we know from \cite{stewart1973} that there exists a unitary matrix $Q$ such that $UQ$ is an invariant subspace of  $f(S^{(r)}+\Delta)$ if $$0\leq 4\beta^2\left\|\Gamma\left[\Delta\right]\right\|_F \leq \Sigma_{k,k}^2-\sigma_{1,1}^2.$$ 
If $\left\|\Delta\right\|_F$ is sufficiently small, the rotation matrix $Q$ will not perturb too much the singular values of $f(S^{(r)})$, so $UQ$ will not only be an invariant but also the dominant subspace of $f(S^{(r)}+\Delta)$, hence the local convergence of the low-rank iterative scheme is guaranteed for sufficiently small $\beta$. This leads to the following bound for the distance between $S^{(r)}$ and the projection of $f(S^{(r)}+\Delta)$
\begin{equation*}
\norm{S^{(r)}-\Pi^{(r)}\left[f(S^{(r)}+\Delta)\right]}_F \leq\gamma\norm{\Delta}_F
\end{equation*}
where $\gamma < 1$ if $$\beta^2 < \frac{1}{\norm{A\otimes A + A^T\otimes A^T}_2\left(\frac{4\norm{\Sigma^2}}{\Sigma_{k,k}^2-\sigma_{1,1}^2}+1\right)}$$
which shows the existence of $\beta$ such that the iteration \eref{eq:LR} converges.

\noindent In the next section, we will apply our low-rank iterative scheme to {Erd\H{o}s-R\'{e}nyi} random graphs and demonstrate that it allows to successfully extract roles in those networks.

\section{Numerical experiments}

\begin{figure*}[t]
	\centering
	\parbox{0.98\textwidth}{
	\centering
\includegraphics[height=0.85\textheight]{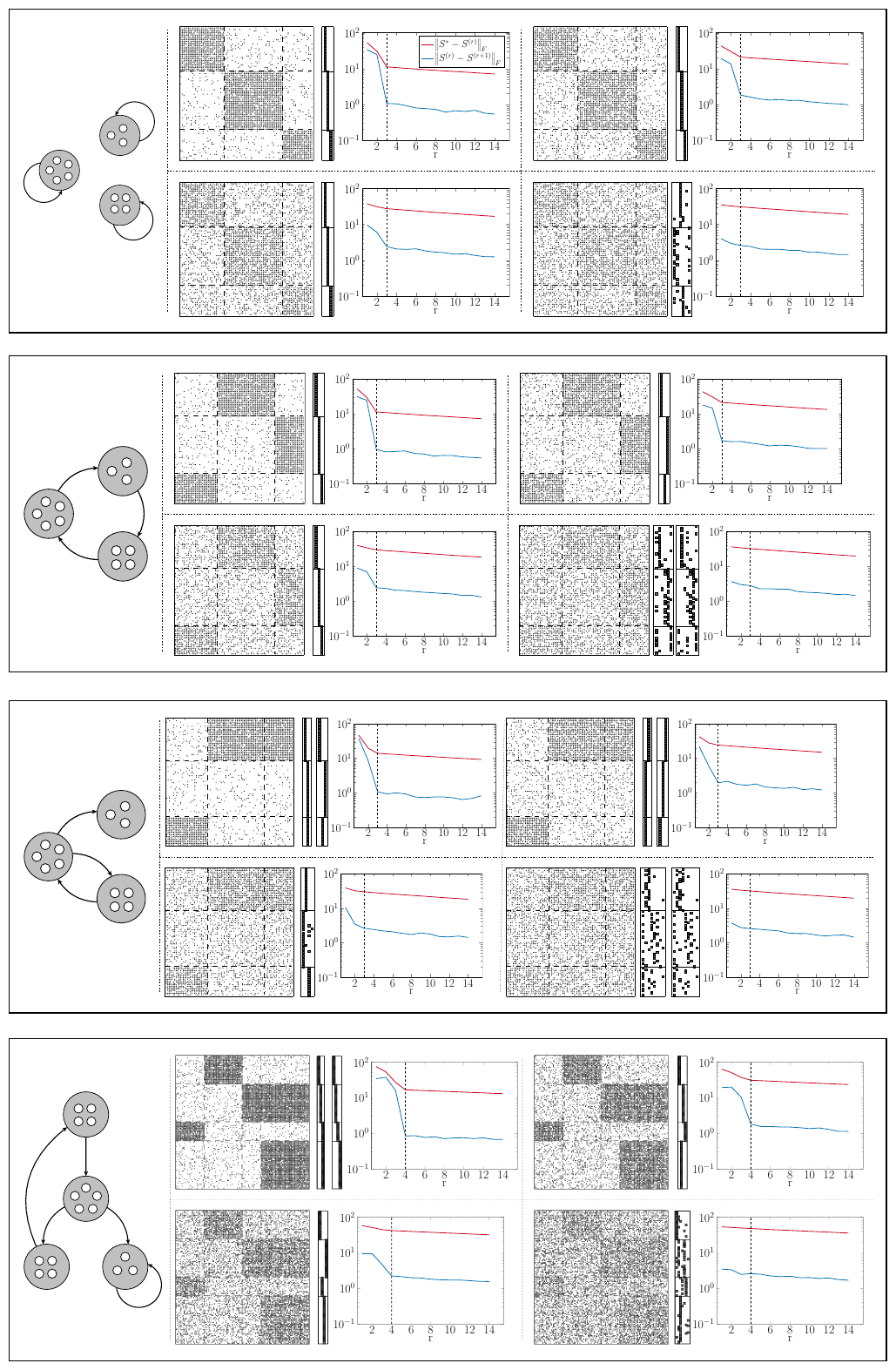}      
	}
    \caption{Evolution of the adjacency matrix, the extracted roles and the low rank similarity measure $S^{(r)}$ for different random graphs. Each panel corresponds to a chosen role graph represented in the first column. Each panel is divided in $4$ sections corresponding to different values of $p_{in}$ and $p_{out}$. In each section, we present one realization of the adjacency matrix, then the extracted role assignment for each node and finally the evolution $\norm{S^*-S^{(r)}}_F$ and $\norm{S^{(r)} - S^{(r+1)}}_F$ for increasing values of $r$.
}
	\label{fig:rank}
\end{figure*}

We applied our similarity measure to extract roles in {Erd\H{o}s-R\'{e}nyi} random graphs containing a block structure. To build such graphs, we first choose a directed role graph $G_B(V_B,E_B)$, \ie each node in  $G_B$ defines a role that we would like to identify. Some of the role graphs that we considered are represented in the first column of each panel of \fref{fig:rank}. As previously, in the role graphs, the large gray filled circles represent the different roles and the small white circles represent the nodes of the graph. The role graph in the first panel corresponds to a community structure where nodes in a role interact mainly with other nodes in the same role. This kind of role graph often occurs when considering human interactions in online social networks for example \cite{Kumpula2009} but has been observed in many other networks \cite{Fortunato2010}. The second panel represents a block cycle role graph, already presented in \fref{fig:cycle}, where each node interacts mainly with nodes in the following role in the cycle. This role graph might represent the behavior of animals in some particular food webs. In the third and fourth panels, the role graphs were simply chosen as representative examples for more complex role interactions without precise real life example in mind.

Once the role graph $G_B$ has been chosen, we build a random graph $G_A(V_A,E_A)$ where each node in $G_A$ has a corresponding role in $G_B$. That is, for each node $i\in V_A$, we select a role $R(i)\in V_B$. Then, we add the edges in $E_A$ using $2$ probability parameters. For every pair of nodes $i,j\in V_A$, we add the edge $(i,j)\in E_A$ with probability $p_{in}$ if there is an edge between the corresponding roles in $G_B$, \ie $\left(R(i),R(j)\right)\in E_B$. If there is no edge between the corresponding roles in $G_B$, the edge is still added with a probability $p_{out}$.
If $p_{in}$ is much larger than $p_{out}$, then the role graph $G_B$ is accurately representing the different roles in the graph $G_A$ and it is expected that the pairwise similarity $S^*$ between the vertices $V_A$ should allow the extraction of those roles. On the other hand, if $p_{out}$ is much larger than $p_{in}$, then the different roles in $G_A$ are more closely represented by the complement graph of $G_B$ represented by the adjacency matrix $\mathbf{1}\mathbf{1}^T-B$. However, the role structure is still strongly existing in this complement graph and it is expected that the similarity measure $S^*$ should still be able to differentiate them. It is when the $2$ probabilities $p_{in}$ and $p_{out}$ are close to each other that extracting the different roles becomes challenging but, at the same time, the graph becomes closer to a uniform {Erd\H{o}s-R\'{e}nyi} graph which is known to be free of any structure.

\begin{figure*}[t]
	\centering
	\parbox{1.95\columnwidth}{
	\centering
\includegraphics[width=1.9\columnwidth]{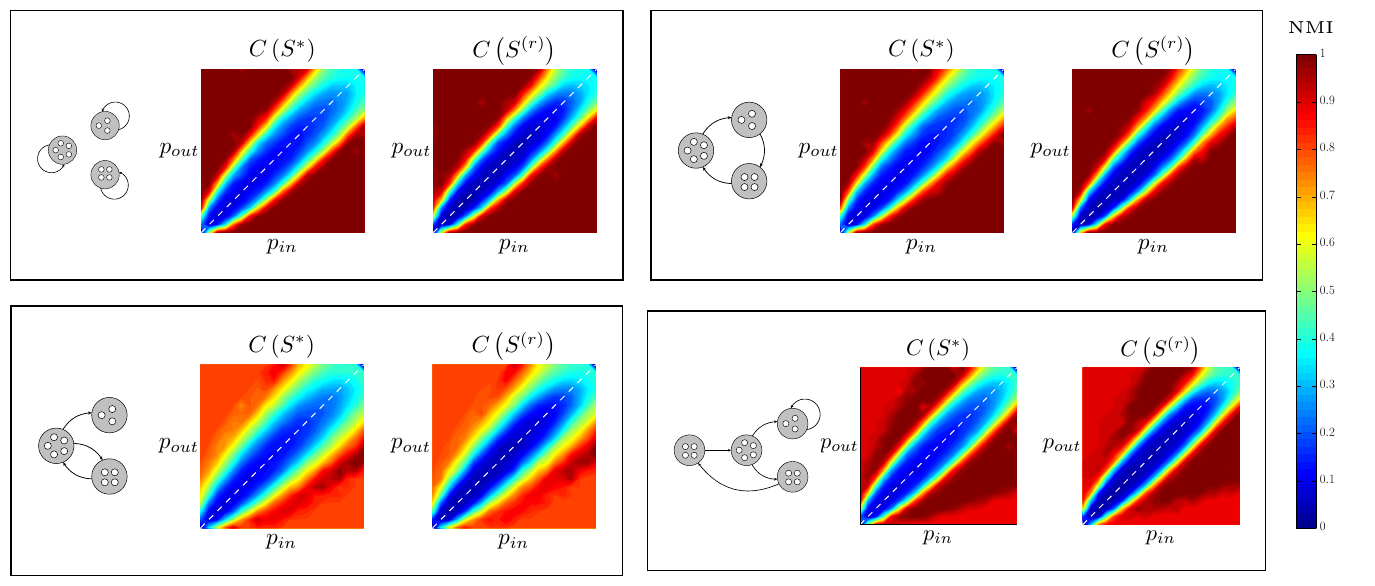}      
	}
    \caption{Average normalized mutual information between the exact role structure and the extracted role structure using the full rank and the low rank similarity measure on different role graphs.
}
	\label{fig:quality}
\end{figure*}

Each of the panels of \fref{fig:rank} is divided into $4$ sections corresponding to different values of $p_{in}$ and $p_{out}$ for a single role graph, as follows
\def\arraystretch{1.5}
$$
\begin{array}{|c|c|c|}
\hline
\multirow{2}{*}{\quad} & p_{in} = 0.9 / p_{out} = 0.1 & p_{in} = 0.8 / p_{out} = 0.2 \\ \cline{2-3}
				& p_{in} = 0.7 / p_{out} = 0.3 & p_{in} = 0.6 / p_{out} = 0.4 \\
				\hline
\end{array}
$$


\noindent In each section of a panel of \fref{fig:rank}, we first present the adjacency matrix of one realization of the random graphs $G_A$ generated. For visual clarity, the matrices have been permuted such that nodes in the same role are next to each other. Then, we represent the role assignment of each node extracted using our low rank similarity matrix $S^{(r)}$ for $r=10$. Based on the similarity measure $S^{(r)}$, we extract the role assignment of each node using the community detection algorithm presented in \cite{Browet2013}. Indeed, within each role, we expect nodes to be highly similar in their neighborhood patterns. This should lead to a similarity graph, whose weighted adjacency matrix is the similarity matrix $S^{(r)}$, with groups of highly connected nodes in each role, hence containing a community structure. Since the algorithm produces hierarchical communities, we present each level of roles when different levels of clustering were extracted from the similarity graph.

\noindent The last plot in each section of a panel in \fref{fig:rank} represents the evolution of $S^{(r)}$ for increasing values of $r$. That is, we compute the norm of the difference between the full rank and each low rank solutions, $\norm{S^*-S^{(r)}}_F$, and between consecutive low rank solutions, $\norm{S^{(r)} - S^{(r+1)}}_F$, for increasing values of $r$. This should reveal the minimal rank required for $S^{(r)}$ to be a qualitatively good approximation of $S^*$.

The results of \fref{fig:rank} clearly show that the different roles within each network can be well extracted using the low rank similarity graph up to some high level of noise. In the first role network, each community is correctly extracted until $p_{in} = 0.6$ and $p_{out} = 0.4$. However, even if the network is really noisy for those parameters, as represented by the adjacency matrix, the first and the third communities are pretty well clustered and the second community is mainly split in $2$. The same observation might be done for the block cycle role graph for which all the roles are perfectly extracted for the first three probability parameters. Again, in the last section, the second and third roles are essentially split in $2$ different clusters but there is only a few nodes with inappropriate role assignments in the final level of clustering. Those first $2$ role graphs have a strong role structure and adding any edge in the role graph would not alter it, \ie this will not create isomorphic roles. This explains why the role structures are correctly extracted even for high level of noise.

The third role graph is less strongly defined because if one edge is added from the second block to the first block, the second and the third roles would become isomorphic. Indeed, we observe that some nodes are incorrectly clustered from the second role to the third role for $p_{in}=0.7$ and $p_{out}=0.3$. This might also explain why the second and third roles are grouped together in the final level of clustering for the first two probability parameters but this might also be due to some resolution limit of the community detection algorithm \cite{Schaub2012,Fortunato2007}. For high level of noise, clustering the pairwise similarity matrix does not provide an accurate result, however, the adjacency matrix clearly indicates that the role structure is very weak.

In the last role graph composed of $4$ distinct blocks, the results are again reasonably good. Except for an additional merge in the last level of clustering for $p_{in}=0.9$ and $p_{out}=0.1$, all the roles are correctly extracted, and even for the last probability parameters, leading to a high level of noise, each role tends to be correctly extracted. There is only a small number of nodes incorrectly classified for the first and last blocks and the second and third blocks are only bisected as previously observed.

We also observe that the evolution of the low rank similarity matrix $S^{(r)}$ might be used to reveal the number of roles in the networks. Indeed, when the different roles in the networks are strongly defined, we observe an abrupt variation in the decay of the norm of the differences $\norm{S^*-S^{(r)}}_F$ and $\norm{S^{(r)}-S^{(r+1)}}_F$. This abrupt variation indicates that we do not need to consider larger values of the rank to extract qualitatively good roles in the networks, since the gain in precision for the similarity measure starts to decrease very slowly afterwards. What is also interesting is that this abrupt variation always occurs when the rank hits the exact number of roles in the networks. When the networks are highly noisy, we do not observe such an abrupt variation which could indicate that the clustering of the nodes according to the similarity matrix will not produce relevant results. Observing the evolution of the low rank similarity matrix could become a strong indicator of the quality of the extracted roles for real networks when the exact block structure is not known.

Finally, we compare quantitatively the extracted clusters using the full rank similarity $S^*$ and the low rank similarity measure $S^{(r)}$. For each of the different role graphs previously introduced, we compute the normalized mutual information (\textit{NMI}) \cite{Danon2005} between the exact role structure and the extracted role assignments using $S^*$ or $S^{(r)}$ and the community detection algorithm. The NMI ranges in $[0,1]$ and is large if the two distributions are similar. More precisely, for each role graph, we generate $20$ random graphs for each couple of probability parameters $p_{in}$ and $p_{out}$ in $[0,1]$ with a discretization step size of $0.05$, and we compute the average NMI on those $20$ realizations of the {Erd\H{o}s-R\'{e}nyi} random graphs. The results are presented in \fref{fig:quality}. As expected, we observe that the extracted roles are accurate when either $p_{in} >> p_{out}$ or the opposite. As we mentioned previously, the third role graph seems harder to recover due to either a resolution limit phenomenon or to the almost isomorphic behavior of two of the role nodes. Nevertheless, we observe that the low rank similarity matrix $S^{(r)}$ produces almost identical results than the full rank similarity $S^*$. This leads us to conclude that, if the rank is sufficiently large, one can always use our low rank pairwise similarity measure to extract role structures in networks. The low rank similarity matrix will always be easier to compute and will produce highly similar results.

\section{Conclusion}

In this paper, we present a pairwise similarity measure between the nodes of a graph that allows the extraction of roles or block structures within the graph. Those roles generalized the concept of communities often studied in the literature. Then, we present a low rank iterative scheme to approximate the pairwise similarity measure and prove its convergence when the parameter $\beta$ is sufficiently small. We applied the similarity measure and its low rank approximation to {Erd\H{o}s-R\'{e}nyi} random graphs containing a block structure and showed that, if the noise level is not too large and the block structure correctly represents the different roles of the nodes in the network, our similarity measure and its low rank approximation accurately extract these blocks. We also show that analyzing the evolution of the low rank similarity measure might reveal the number of roles in the networks and also might indicate if the extracted cluster are relevant. Finally, we demonstrate that the pairwise similarity measure and the low rank approximation produce very similar results, hence justifying the use of the low rank approximation in practical examples when computing the full rank measure is computationally too expensive. In future works, we plan to apply our low rank similarity measure to other kinds of random graphs, \eg scale-free networks. We will also apply our measure to real networks like food webs, international exchange networks or words graphs to automatically uncover similar type of words in the construction of sentences, known as ``tagging'' in natural language processing. We will also analyze the behavior of our similarity measure against weighted networks.


\addtolength{\textheight}{-12cm}   



%

\section*{ACKNOWLEDGMENT}

We would like to thank N. Boumal, R. Jungers and J. Hendrickx for their useful comments regarding the subject.\\

\noindent This paper presents research results of the Belgian Network DYSCO (Dynamical Systems, Control, and Optimization), funded by the Interuniversity Attraction Poles Programme, initiated by the Belgian State, Science Policy Office. The scientific responsibility rests with its author(s).


\bibliographystyle{IEEEtran}
\bibliography{IEEEabrv,RoleCitation}

\begin{thebibliography}{10}
\providecommand{\url}[1]{#1}
\csname url@rmstyle\endcsname
\providecommand{\newblock}{\relax}
\providecommand{\bibinfo}[2]{#2}
\providecommand\BIBentrySTDinterwordspacing{\spaceskip=0pt\relax}
\providecommand\BIBentryALTinterwordstretchfactor{4}
\providecommand\BIBentryALTinterwordspacing{\spaceskip=\fontdimen2\font plus
\BIBentryALTinterwordstretchfactor\fontdimen3\font minus
  \fontdimen4\font\relax}
\providecommand\BIBforeignlanguage[2]{{%
\expandafter\ifx\csname l@#1\endcsname\relax
\typeout{** WARNING: IEEEtran.bst: No hyphenation pattern has been}%
\typeout{** loaded for the language `#1'. Using the pattern for}%
\typeout{** the default language instead.}%
\else
\language=\csname l@#1\endcsname
\fi
#2}}

\bibitem{Arenas2008}
\BIBentryALTinterwordspacing
A.~Arenas, A.~Fernandez, and S.~Gomez, ``{Analysis of the structure of complex
  networks at different resolution levels},'' \emph{New Journal of Physics},
  vol.~10, no.~5, p.~23, 2007. [Online]. Available:
  \url{http://arxiv.org/abs/physics/0703218}
\BIBentrySTDinterwordspacing

\bibitem{Porter2009}
\BIBentryALTinterwordspacing
M.~A. Porter, J.-P. Onnela, and P.~J. Mucha, ``{Communities in networks},''
  \emph{Notices of the AMS}, vol.~56, no.~9, pp. 1082--1097, 2009. [Online].
  Available: \url{http://www.ams.org/notices/200909/rtx090901082p.pdf}
\BIBentrySTDinterwordspacing

\bibitem{Fortunato2010}
\BIBentryALTinterwordspacing
S.~Fortunato, ``{Community detection in graphs},'' \emph{Physics Reports}, vol.
  486, no. 3-5, pp. 75--174, 2010. [Online]. Available:
  \url{http://linkinghub.elsevier.com/retrieve/pii/S0370157309002841}
\BIBentrySTDinterwordspacing

\bibitem{Wasserman1994}
\BIBentryALTinterwordspacing
S.~Wasserman and K.~Faust, \emph{{Social Network Analysis: Methods and
  Applications}}.\hskip 1em plus 0.5em minus 0.4em\relax Cambridge University
  Press, 1994. [Online]. Available: \url{http://www.amazon.com/dp/0521387078}
\BIBentrySTDinterwordspacing

\bibitem{Cason2012}
T.~P. Cason, ``Role extraction in networks,'' Ph.D. dissertation, Universite
  catholique de Louvain, 2012.

\bibitem{reichardt2007}
J.~Reichardt and D.~R. White, ``{Role models for complex networks},''
  \emph{Eur. Phys. J. B}, vol.~60, no.~2, pp. 217--224, 2007.

\bibitem{reichardt2006}
\BIBentryALTinterwordspacing
J.~Reichardt and S.~Bornholdt, ``{Statistical mechanics of community
  detection},'' \emph{Physical Review E}, vol.~74, no.~1, p. 016110, 2006.
  [Online]. Available: \url{http://link.aps.org/doi/10.1103/PhysRevE.74.016110}
\BIBentrySTDinterwordspacing

\bibitem{Denayer2012}
D.~Denayer, ``Mod\'{e}lisation par r\^oles de grands graphes,'' Master's
  thesis, Universite catholique de Louvain, 2012.

\bibitem{Cooper2011}
K.~{Cooper} and M.~{Barahona}, ``{Role-similarity based comparison of directed
  networks},'' \emph{ArXiv e-prints}, Mar. 2011.

\bibitem{Diaz2013}
M.~Beguerisse-Diaz, B.~Vangelov, and M.~Barahona, ``Finding role communities in
  directed networks using role-based similarity, markov stability and the
  relaxed minimum spanning tree,'' \emph{CoRR}, vol. abs/1309.1795, 2013.

\bibitem{Blondel2004}
\BIBentryALTinterwordspacing
V.~D. Blondel, A.~Gajardo, M.~Heymans, P.~Senellart, and P.~Van~Dooren, ``{A
  Measure of Similarity between Graph Vertices: Applications to Synonym
  Extraction and Web Searching},'' \emph{SIAM Review}, vol.~46, no.~4, pp.
  647--666, 2004. [Online]. Available: \url{http://dx.doi.org/10.2307/20453570}
\BIBentrySTDinterwordspacing

\bibitem{stewart1973}
\BIBentryALTinterwordspacing
G.~W. Stewart, ``Error and perturbation bounds for subspaces associated with
  certain eigenvalue problems,'' \emph{SIAM Review}, vol.~15, no.~4, pp. pp.
  727--764, Oct. 1973. [Online]. Available:
  \url{http://www.jstor.org/stable/2028728}
\BIBentrySTDinterwordspacing

\bibitem{Kumpula2009}
\BIBentryALTinterwordspacing
J.~Kumpula, J.-P. Onnela, J.~Saram\"{a}ki, J.~Kert\'{e}sz, and K.~Kaski,
  ``{Model of community emergence in weighted social networks},''
  \emph{Computer Physics Communications}, vol. 180, no.~4, pp. 517--522, Apr.
  2009. [Online]. Available:
  \url{http://linkinghub.elsevier.com/retrieve/pii/S0010465508004347}
\BIBentrySTDinterwordspacing

\bibitem{Browet2013}
A.~{Browet}, P.-A. {Absil}, and P.~{Van Dooren}, ``{Fast community detection
  using local neighbourhood search},'' \emph{ArXiv e-prints}, Aug. 2013.

\bibitem{Schaub2012}
\BIBentryALTinterwordspacing
M.~T. Schaub, J.-C. Delvenne, S.~N. Yaliraki, and M.~Barahona, ``{Markov
  dynamics as a zooming lens for multiscale community detection: non
  clique-like communities and the field-of-view limit},'' \emph{PLoS ONE}, pp.
  1--11, 2012. [Online]. Available:
  \url{http://www.plosone.org/article/info:doi/10.1371/journal.pone.0032210}
\BIBentrySTDinterwordspacing

\bibitem{Fortunato2007}
\BIBentryALTinterwordspacing
S.~Fortunato and M.~Barth\'{e}lemy, ``{Resolution limit in community
  detection.}'' \emph{Proceedings of the National Academy of Sciences of the
  United States of America}, vol. 104, no.~1, pp. 36--41, Jan. 2007. [Online].
  Available: \url{http://www.pnas.org/cgi/content/abstract/104/1/36}
\BIBentrySTDinterwordspacing

\bibitem{Danon2005}
\BIBentryALTinterwordspacing
L.~Danon, J.~Duch, A.~Diaz-Guilera, and A.~Arenas, ``{Comparing community
  structure identification},'' \emph{Journal of Statistical Mechanics: Theory
  and Experiment}, vol. 2005, no.~09, p.~10, 2005. [Online]. Available:
  \url{http://arxiv.org/abs/cond-mat/0505245}
\BIBentrySTDinterwordspacing

\end{thebibliography}

\end{document}